\documentclass[letterpaper]{jpconf}
\usepackage{graphicx}
\usepackage{graphicx}
\usepackage{fancyhdr}
\usepackage{bm}        
\usepackage{amssymb}   
\usepackage{amsmath}   
\usepackage{xspace}
\def\D0{D\O}                            			
\newcommand{\Ptg}{p_{T}^{\gamma}}				
\newcommand{\rJLIP}{\ensuremath{P_{b\text{-jet}}}\xspace}

\newcommand{\gb}{\ensuremath{\gamma+{b}}\xspace}
\newcommand{\gc}{\ensuremath{\gamma+{c}}\xspace}

\begin{document}
\title{
	Measurement of $\gamma~+~c~+$~X
	and $\gamma~+~b~+$~X Production
	Cross Sections at $\sqrt{s} = 1.96$ TeV
       }

\author{Daniel Duggan}

\address{Florida State University}

\begin{abstract}
The photon plus heavy-flavour quark ($b, c$) final state provides
a unique and valuable window into both the sea quark content of
the proton and the splitting of gluons into heavy-flavour quark
pairs. A new combination of experimental techniques has provided
the basis for the first measurements of the differential
$\gamma + c +$ X and $\gamma + b +$ X production cross sections at
$\sqrt{s} = 1.96$ TeV. The measurements use $\sim$1 fb$^{-1}$ of data
from $p\bar{p}$ collisions collected with the \D0 detector.
\end{abstract}
\section{Introduction}

The results presented are the first measurements of the differential
$\gamma + c$ jet and $\gamma + b$ jet production cross sections at
a hadron-hadron collider~\cite{PRL}. The measurements used $\sim1$~fb$^{-1}$ 
of data and were performed at the Fermilab Tevatron $p\bar{p}$
collider at $\sqrt{s}$ = 1.96 TeV collected using the \D0 detector. 
The presented production cross sections are differential 
with respect to the photon candidate's transverse momentum ($\Ptg$)
and are divided into five $\Ptg$ bins. The measurements are also
differential with respect to the leading jet's rapidity ($y^\text{jet}$)
and the leading photon's rapidity ($y^{\gamma}$) and are binned
in two regions of photon--\,jet rapidities:
$y^\gamma \cdot y^\text{jet} > 0$
and $y^\text{jet} \cdot y^{\gamma} < 0$.

\section{Data Selection}
Events are recorded using the \D0 detector~\cite{D0_det} and must
contain at least one photon candidate and at least one heavy-flavour
jet candidate. The photon and jet candidates with the highest
transverse momenta (leading) are selected. The $p_{T}$ of the photon
candidate must be greater than 30 GeV/$c$ with $|y^{\gamma}| < 1.0$.
The leading jet must satisfy $p_{T} > 15$ GeV/$c$ and
$|y^\text{jet}| < 0.8$. The primary collision vertex
must be located within 35~cm of the detector's center along the
beam pipe, reconstructed with at least three associated tracks. 
To suppress background events coming from cosmic-ray muons and
$W\rightarrow \ell\nu$ decays, the total missing transverse energy
($E_{T}^{Miss}$) in the event must satisfy the condition of
 $E_{T}^{Miss} < 0.7 \cdot \Ptg$.

This analysis employs triggers that distinguish large depositions
of energy in the electromagnetic (EM) calorimeter, and events are
selected if at least one of these triggers has fired. The photon 
candidate is reconstructed, forming a cluster of energy. The
reconstruction procedure sums the total energy deposited in the
calorimeter within a cone of radius ${\cal{R}} = 0.2$, where
${\cal{R}} = \sqrt{(\Delta\eta)^2 +(\Delta\phi)^2}$. At least 96\%
of the photon candidate's energy must be in the EM calorimeter,
and it must be spatially well isolated. To be considered isolated,
the total energy in a cone of ${\cal R} = 0.4$, excluding the candidate's
energy in the EM calorimeter ($E_{\text{EM}}$), must be less than
7\% of $E_{\text{EM}}$. Additionally, the energy deposition within the
third layer of the EM calorimeter must be consistent with that of
an electromagnetic shower. Photons traversing the \D0 detector
typically do not leave tracks, thus the photon candidate is
required to have less than a 0.1\% probability of being spatially
matched to any track in the event. To reject additional backgrounds
coming from dijet events, an an artificial neural network
($\gamma-$ANN)~\cite{gamjet_PLB} is used with the requirement of 
$\gamma-$ANN $> 0.7$. 

Jets are reconstructed using \D0's RunII algorithm~\cite{c:Run2Cone}
with a radius 0.5. The leading jet is subjected to the additional 
requirement of having at least two associated tracks with hits in
the silicon tracker, one with $p_T > 1.0$ GeV/$c$ and the another with
$p_T > 0.5$ GeV/$c$. Heavy-flavour jets are identified using information
arising from the long lifetimes of their hadrons. To best enhance
the fraction of jets originating from heavy-flavour quarks, an
artificial neural network ($b-$ANN) is employed. The 
$b-$ANN~\cite{c:bNN} is trained such that the output of heavy-flavour
jets tends toward one and that of light ($u,~d,~s$ quarks and gluons)
jets tends to zero. For this selection, the requirement of
$b-$ANN $> 0.85$ must be satisfied.

\section{Purity Estimates}
The fraction of direct photons in the final data sample is determined
from a simultaneous fit of the $\gamma-$ANN output shape in data using
templates derived from Monte Carlo (MC) signal photon and background dijet 
events simulated using {\sc pythia}~\cite{PYT}. The fitting procedure
determines the linear combination of the signal and background templates
that best describes the output shape in data. The fit is performed in 
each $\Ptg$ bin for both rapidity regions separately.

\begin{figure}
\centering
\includegraphics[width=0.55\linewidth]{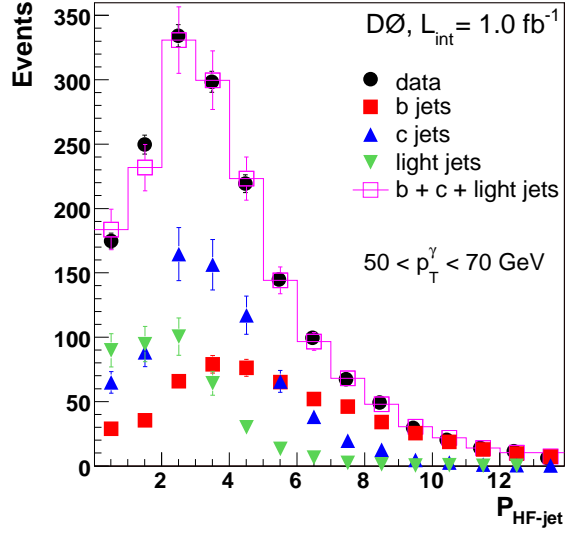}
\caption{Distribution of the \rJLIP discriminant in the final data
	 sample in the range of $50 < \Ptg < 70$ GeV/$c$. The 
	 templates for light, $c$ and $b$ jets have been normalized
	 to the number of events in data and then weighted by their
	 found flavour fractions, and the shape of their sum is found
	 to be in good agreement with that of the data. The shown 
	 uncertainties for the data are statistical in nature, whereas
	 for the templates they incorporate statistical and fitting
	 uncertainties.}
\label{fig:bNNfit}
\end{figure}

The fraction of $c$ and $b$ jets in the final data sample is determined
using a similar method to that of photon purity technique. However, for
the heavy-flavour fraction fitting technique, another discriminant, 
$\rJLIP$ was used. $\rJLIP$ is defined as
$\rJLIP=-\ln\prod_{i}{P_{\rm track}^{i}}$, where $P_{\rm track}^{i}$ is
the probability of a track in the jet cone to originate from the primary
collision vertex, omitting the least likely track to have come from this
vertex. The value of \rJLIP is large for $b$ jets, and tends toward zero
for light jets. The shape of the \rJLIP distribution in data is fit from
MC templates for both $b$ jets and $c$ jets, with the light jet template
coming from a light jet enriched sample in data. The fit is performed in
each $\Ptg$ bin separately, and the quality of the fit in each bin is of
good quality. As an example, the result of the fit for $50 < \Ptg < 70$ GeV/$c$
is shown in Fig.~\ref{fig:bNNfit}.

\section{Cross Section Results}
\begin{figure}
\centering
\includegraphics[width=0.55\linewidth]{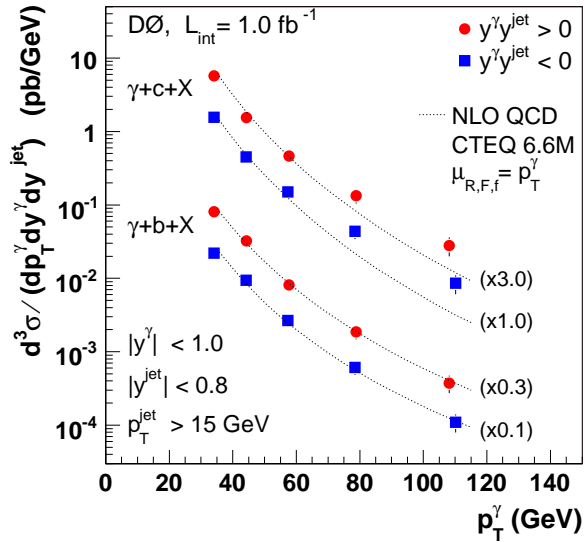}
\caption{The \gb and \gc differential cross sections as a function
	 of $\Ptg$ for both rapidity regions. The data points include
	 the overall uncertainties from the measurement, and the 
	 theoretical predictions are displayed as dotted lines.}
\label{fig:xsect}
\end{figure}

The measured differential cross sections are presented in five $\Ptg$ bins
and two regions of $y^\gamma \cdot y^\text{jet}$, and can be seen in
Fig.~\ref{fig:xsect} for both the \gb and \gc processes. Due to the finite
resolution of the calorimeter, $\Ptg$ smearing effects were accounted
for using the unfolding method described in Ref.~\cite{D0_unsmearing}.
Statistical uncertainties for these results vary between 0.2\% for 
$30 < \Ptg < 40$ GeV/$c$ to $\sim$9\% for $90 < \Ptg < 150$ GeV/$c$, and
the systematic uncertainties range from 15\% to 28\%. The theoretical
predictions from next-to-leading (NLO) order calculations~\cite{Tzvet}
are also presented in Fig.~\ref{fig:xsect}, where the renormalization
scale $\mu_{R}$, factorization scale $\mu_{F}$ and fragmentation scale
$\mu_f$ are all set equal to $\Ptg$.

\begin{figure}
\centering
\includegraphics[width=0.55\linewidth]{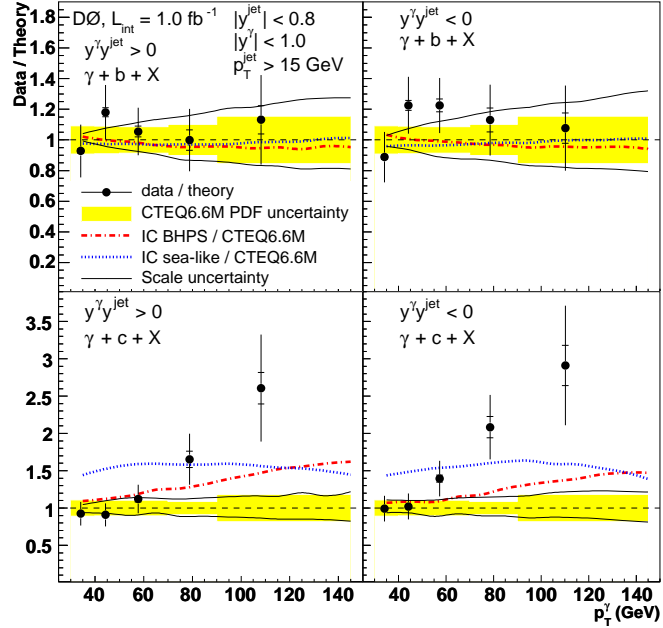}
\caption{The ratios of the measured to the predicted cross sections
	 for \gb production (top) and \gc production (bottom) in 
	 each rapidity region. The data points include both their
	 statistical (inner tick) and overall uncertainties 
	 (entire error bar). The uncertainties from the theoretical 
	 predictions include those from the {\sc cteq} 6.6M 
	 PDFs (yellow band) and from the choice of scale (full line).
	 The ratio of two intrinsic charm models to the standard
	 theoretical predictions are also included (dashed lines).}
\label{fig:xsectratio}
\end{figure}

The ratios of the measured to the predicted cross sections for both \gb 
and \gc cross sections in the two rapidity regions are shown in 
Fig.~\ref{fig:xsectratio}. The uncertainty due to the scale choice is
estimated by simultaneously varying all three scales by a factor of two,
$\mu_{R,F,f}=0.5 p_T^\gamma$ and $\mu_{R,F,f}=2 p_T^\gamma$.
The {\sc cteq}6.6M parton distribution functions (PDFs) were used in
the theoretical calculations, and their corresponding uncertainties
were calculated according to the prescription in Ref.~\cite{CTEQ}. 
Agreement within uncertainties can be seen in Fig.~\ref{fig:xsectratio}
for the \gb cross sections between data and theory in both rapidity
regions. The \gc cross sections show agreement within uncertainties
up to $\Ptg \sim 70$ GeV/$c$; however, there is a rising discrepancy
between data and theory that grows as $\Ptg$ increases. Certain models,
including those that include an intrinsic charm (IC) component to the
proton, predict higher cross sections for \gc production. Although at
least one IC model shows better agreement with the data than the standard
theoretical predictions do, neither model shows agreement for the entire
$\Ptg$ range in either rapidity region.

\end{document}